\documentstyle[aps,preprint]{revtex}
\tightenlines
\begin{document}
\draft
\title{Gauge vortex dynamics at finite mass of bosonic fields.}
\author{A.A. Kozhevnikov}
\address{Laboratory of Theoretical Physics,\\
S.L. Sobolev Institute for Mathematics,\\
630090, Novosibirsk-90, Russian Federation
\footnote{Electronic address: kozhev@math.nsc.ru}
}
\date{\today}
\maketitle
\begin{abstract}
The simple derivation of the string equation of motion adopted in the 
nonrelativistic case is presented, paying the special attention to the
effects of finite masses of bosonic fields of an Abelian Higgs
model. The role of the finite mass effects in the evaluation of various
topological characteristics of the closed strings is discussed. 
The rate of the dissipationless helicity change is calculated.
It is demonstrated how the conservation of the sum of the twisting and
writhing numbers of the string is recovered despite the changing helicity.
\end{abstract}

\narrowtext

Stringlike defects are widely discussed as the possible remnants
surviving the epoch of  phase transitions in the early Universe
\cite{vilenkin,hindmarsh}. The dynamics underlying the evolution of the
cosmic string network is usually assumed to be governed by the Nambu-Goto
(NG) action \cite{vilenkin,hindmarsh,foerst}.
Then a typical velocity of the string segment is of the order of that of
light. The derivations of the string equation of motion from the 
field theoretic lagrangian existing in the literature are focused on
an expansion in the parameter called the core radius that is 
proportional to the inverse mass of the
Higgs field. However, in the London limit to be specified below, the 
mass of the vector field of the gauge model is much smaller than the 
Higgs boson mass, so that there is another length parameter much greater
than the core radius. It raises the natural question of to what
extent the finite vector mass will influence the evaluation of 
various characteristics of the string that depend essentially on the
vector field configuration. 

In the present paper we address this question and present the derivation
of the string equation of motion upon keeping the trace of the finite
mass of the vector field. The role of the finite mass in the evaluation 
of such
topological characteristics of the closed string as the linking and
twisting numbers is discussed. Related to them is 
helicity of the mirror-noninvariant string configuration.
Its change in the course of the contour evolution will also be discussed,
with the special attention paid to the role of the effects of the
finite mass of
the vector boson. As is known \cite{cornwall,giovan}, the dynamics of such
configurations, through the anomaly equation,
may be important for the dynamics of some fermionic charges.
It will be shown how conservation of the sum of the twisting and writhing
numbers of the string contour is recovered in the case of the changing
helicity.

We start with the Abelian Abrikosov-Nielsen-Olesen
(ANO) string \cite{ano} in the Higgs model with the action
\begin{equation}
S=\int d^4x\left[-{1\over4}F^2_{\mu\nu}+|(\partial_\mu+igA_\mu)\phi|^2
\right.
-\left.{\lambda^2\over2}\left(|\phi|^2-{\eta^2\over2}\right)^2\right],
\label{action}
\end{equation}
where $F_{\mu\nu}=\partial_\mu A_\nu-\partial_\nu A_\mu$,
 and obtain the  closed form of the string action in the London limit.
This is the limit of $m_H\gg m_V$; $\ln m_H/m_V$ is also large, where
$m_V=g\eta/2$ and $m_H=\lambda\eta$ are the
masses of the gauge and Higgs bosons, respectively, and  $\lambda$ and
$\eta/\sqrt{2}$
are the Higgs field self-coupling and magnitude.
The nonstationary field configuration of the gauge
string is expressed through  the
spacetime dependent phase $\chi\equiv\chi({\bf x},t)$ of the
Higgs field $\phi({\bf x},t)=\eta\exp(i\chi)/\sqrt{2}$.
As is known, it is the dynamics of the phase $\chi$ of the scalar
field, not of its radial part (modulus), that is  essential in
the London limit.
One can ignore the details of the Higgs field profile, taking it to be
uniform $\eta/\sqrt{2}$ in all coordinate space except the vortex line where
it approaches zero at the characteristic distances $\sim m_H^{-1}$.

The equation for the magnetic field is obtained upon varying the action
(\ref{action}). It looks as
\begin{equation}
\mbox{\boldmath$\nabla$}\times{\bf H}=\frac{m^2_V}{g}
\mbox{\boldmath$\nabla$}\chi-m^2_V{\bf A},
\label{eq22}
\end{equation}
and can be solved in the momentum representation to give the magnetic field
strength
\begin{equation}
{\bf H}({\bf k},t)=\frac{2\pi}{g}\cdot
\frac{m^2_V}{{\bf k}^2+m^2_V}\oint d\sigma{\bf X}^\prime_a
\exp(-i{\bf k\cdot X}_a)
\label{eq23}
\end{equation}
and the vector potential
\begin{equation}
{\bf A}({\bf k},t)=\frac{2\pi}{g}
\Biggl(\frac{1}{{\bf k}^2}-\frac{1}{{\bf k}^2+m^2_V}\Biggr)
\oint d\sigma i[{\bf k}\times{\bf X}^\prime_a]
\exp(-i{\bf k\cdot X}_a).
\label{eq23a}
\end{equation}
The integral over $\sigma$ comes from the equation for the phase $\chi$
read off from Refs.~\cite{lee,orland}, with the proper
continuation to  Minkowski spacetime:
\begin{equation}
\mbox{\boldmath$\nabla$}\times\mbox{\boldmath$\nabla$}
\chi({\bf x},t)=2\pi\oint d\sigma
{\bf X}_a^{\prime}\delta^{(3)}
[{\bf x}-{\bf X}_a(\sigma,t)],
\label{eq8}                                                    %(8)
\end{equation}
where ${\bf X}_a\equiv{\bf X}_a(\sigma,t)$ is the evolving closed
string contour $a$ parametrized by the arclength $\sigma$.
 Hereafter the prime over ${\bf X}$ will
denote a derivative with respect to the corresponding parameter along the
contour, while
the overdot will do the time derivative. The case of many contours is embraced
by taking the sum over individual  contributions on the right hand side of
Eq.~(\ref{eq8}).
The winding number $n$ of the scalar field  is
related to the magnetic flux via the condition of the vanishing
covariant derivative of the Higgs field deep inside in the Higgs condensate,
\begin{equation}
\oint{\bf A}\cdot d{\bf l}={1\over g}\oint
\mbox{\boldmath$\nabla$}\chi\cdot d{\bf l}={2\pi n\over g}\equiv\Phi_0n.
\label{match}
\end{equation}
Hereafter $n$ is taken to be unity.

To specify the dynamical part of the problem, one should write down the
electric field strength
${\bf E}=-\mbox{\boldmath$\nabla$}A_t-\partial_t{\bf A}$,
where
\begin{equation}
A_t=-\frac{m^2_V}{g}(-\mbox{\boldmath$\nabla$}^2+m^2_V)^{-1}
\partial_t\chi
\label{eq26}
\end{equation}
guarantees  finite energy per unit length for the vortex in the nonstatic
situation and replaces the  condition $A_t=0$ appropriate in the static case.
One has
\begin{equation}
{\bf E}={m^2_V\over g}\left(-\mbox{\boldmath$\nabla$}^2+m^2_V\right)^{-1}
(\mbox{\boldmath$\nabla$}\partial_t
-\partial_t\mbox{\boldmath$\nabla$})\chi.
\label{eq27}
\end{equation}
The commutator of the derivatives  is nonzero in view of the
singular character of the phase $\chi$ \cite{lee,orland};
so the Fourier component of ${\bf E}$ becomes
\begin{equation}
{\bf E}({\bf k},t)=-{2\pi\over g}\frac{m^2_V}{{\bf k}^2+m^2_V}\oint d\sigma
(\dot{\bf X}_a\times{\bf X}_a^{\prime})\exp
[-i{\bf k}\cdot{\bf X}_a(\sigma,t)].
\label{dop1}
\end{equation}
Note that ${\bf E}$ looks like the local, at given $\sigma$, boost
of ${\bf H}$.
The Fourier component of
${\bf v}({\bf x},t)\equiv(1/g)\mbox{\boldmath$\nabla$}\chi$
found from Eq. (\ref{eq8}), is
\begin{equation}
{\bf v}({\bf k},t)=\frac{2\pi}{g{\bf k}^2}
\oint d\sigma i[{\bf k}\times{\bf X}^\prime_a]
\exp(-i{\bf k\cdot X}_a),
\label{eq25}
\end{equation}
and the time component can be obtained from the above expression by the local 
boost. Note that the vector potential ${\bf A}$ and the magnetic
field strength ${\bf H}$  can also be expressed through the gradient of the
singular phase $\chi$  as
\begin{eqnarray}
{\bf A}({\bf k},t)&=&
\left(1-\frac{{\bf k}^2}{{\bf k}^2+m^2_V}\right){\bf v}({\bf k},t),
      \nonumber\\
{\bf H}({\bf k},t)&=&{m_V^2\over{\bf k}^2+m^2_V}
i[{\bf k}\times{\bf v}({\bf k},t)].
\label{fields}
\end{eqnarray}
In what follows we will neglect both the close encounters of the
segments of different strings and the segments of the same string
that are labelled by distinct values of the arclength $\sigma$.
Important as they are in the processes of string reconnections, they cannot
be described in the framework of the London approximation.
Substituting Eqs. (\ref{eq23}), (\ref{eq23a}) and (\ref{dop1}) into
Eq. (\ref{action}) one obtains, with the help of the relation
$$\int d^3x{\bf H}^2({\bf x})=\int d^3k|{\bf H}({\bf k})|^2
/(2\pi)^3$$ 
the expression for the action of the single gauge vortex:
\begin{eqnarray}
S_{\rm vortex}&=&{1\over2}
m^2_V\Phi_0^2\int{d^3k\over(2\pi)^3}
{{\bf k}^2\over({\bf k}^2+m^2_V)^2}
\int dt\oint d\sigma_1d\sigma_2\exp\{i{\bf k}\cdot[{\bf X}(\sigma_1)
-{\bf X}(\sigma_2)]\}           \nonumber\\
&& \times\{-{\bf X}^\prime(\sigma_1)\cdot{\bf X}^\prime(\sigma_2)
+[\dot{\bf X}(\sigma_1)\times{\bf X}^\prime(\sigma_1)]\cdot
[\dot{\bf X}(\sigma_2)\times{\bf X}^\prime(\sigma_2)]\}. 
\label{lagr}
\end{eqnarray}
Let us show with a method similar to that of
Refs. \cite{orland} and \cite{sato95} how the known Nambu-Goto form of the
action results from Eq. (\ref{lagr}). To this end one should set the
mass of the gauge boson to infinity,  $m_V\to\infty$,
before the momentum integration. Then
the action becomes, in the gauge $X^0\equiv t=\tau$,
\begin{eqnarray}
S_{\rm NG}&=&{\Phi^2_0\over2}\int d^2s_1d^2s_2\delta^{(4)}[X(s_1)-
X(s_2)]\{-{\bf X^\prime}(s_1)\cdot{\bf X^\prime}(s_2)
\nonumber\\
& &+ [{\bf\dot X}(s_1)\times{\bf X^\prime}(s_1)]\cdot
[{\bf\dot X}(s_2)\times{\bf X^\prime}(s_2)]\},
\label{ng1}
\end{eqnarray}
where  $s_{1,2}\equiv s^A_{1,2}=(\tau_{1,2},\sigma_{1,2})$ is the two-
dimensional vector. Using the Gaussian regularization
of the $\delta$ function and the expansion
\begin{equation}
{\bf X}(s_2)\simeq{\bf X}(s_1)+(s_2-s_1)^A\partial_A
{\bf X}/1!+(s_2-s_1)^A(s_2-s_1)^B\partial_A\partial_B{\bf X}/2!+\cdots
\label{eqcl}
\end{equation}
\cite {sato95,fn1}, valid under the condition
$|{\bf X^{\prime\prime}}(\sigma)|\ll m_V$, one obtains
\begin{eqnarray}
S_{\rm NG}&=&{1\over2}\Biggl({\Phi_0\over2\pi\Lambda^2}\Biggr)^2
\int d^2s_1d^2z\exp(-{1\over2\Lambda^2}z^Az^B\partial_AX^\mu
\partial_BX_\mu)(-{\bf X^\prime}^2+[{\bf\dot X}\times{\bf X^\prime}]^2)
             \nonumber\\
& &={\Phi^2_0\over4\pi\Lambda^2}\int d^2s\sqrt{\mbox{det}
\partial_AX^\mu\partial_BX_\mu},
\label{ng}
\end{eqnarray}
where $\Lambda^{-1}\to\infty$ is an ultraviolet cutoff, $\partial_A=
\partial/\partial z^A$, and $\mbox{det}\partial_AX^\mu\partial_BX_\mu=
-{\bf X^\prime}^2+[{\bf\dot X}\times{\bf X^\prime}]^2$ in the chosen gauge.
Up to an overall factor, the last equality in Eq. (\ref{ng}) is recognized
to be the NG action.

Coming back to the case of large but finite $m_V$, one should first  make
the integration over momenta neglecting  exponentially small terms
and taking into account the fact that
only  nearby segments of the string contour give an appreciable
contribution to the integral over the arclength. One obtains
the action of a single vortex in the form
\begin{equation}
S_{\rm vortex}=
{\pi\over g^2}\ln{m_H\over m_V}\int dt\oint d\sigma
\left\{-m^2_V{\bf X}^{\prime2}
+m^2_V[{\bf\dot X}\times{\bf X}^\prime]^2\right\}.
\label{acti}
\end{equation}
The energy of the electric and magnetic fields is not enhanced
in the London limit and by this reason it is neglected. The natural 
parametrization of the arclength $|{\bf X}^\prime|=1$ is understood.
Here the mass of the Higgs boson $m_H$ appears as the natural
upper limit of the integration over momentum.

In order to be self-consistent, one should convince oneself that
there are stages of the physical motions of the string characterized by the
nonrelativistic speed. To this end one should consider
the circular string loop and find its time to shrink. The
equation of the contour is
\begin{equation}
{\bf X}(\sigma,t)=a\left({\bf e}_x\cos{\sigma\over a}+
{\bf e}_y\sin{\sigma\over a}\right),
\label{eq31}
\end{equation}
where $a\equiv a(t)$ is the time dependent loop radius, ${\bf e}_{x,y}$
being the  unit vector in the corresponding direction. Then the Lagrangian
obtained from the Eq. (\ref{acti}) becomes
\begin{equation}
L=2\pi a\varepsilon_{\rm v}\left(-1+\dot a^2\right),
\label{eq32}
\end{equation}
with $$\varepsilon_{\rm v}={2\pi m_V^2\over g^2}\ln{m_H\over m_V}$$ being
the energy per unit length. The equation of motion, with  initial
conditions in the form $\dot a(0)=0$ and $a(0)=R$, is solved through the
equation of the energy conservation in the process of  collapse,
$$E=2\pi R\varepsilon_{\rm v}=2\pi a\varepsilon_{\rm v}
\left(1+\dot a^2\right).$$ One
obtains an equation determining implicitly the dependence of the loop radius
$a$ on time,
\begin{equation}
R\left({\pi\over2}-\arcsin\sqrt{{a\over R}}\right)+\sqrt{a(R-a)}=t.
\label{eq33}
\end{equation}
One can see from Eq. (\ref{eq33}) that at times $t\ll R$ the velocity of the 
transverse motion is nonrelativistic.

On the other hand, by  direct application of the Hamiltonian formalism
to the action  (\ref{acti}) subjected to the constrains
${\bf X}^{\prime2}=1$ and ${\bf X}^{\prime}\cdot\dot{\bf X}=0$,
one can show that the string stabilized by
some means (say, by  rotation in the plane of the loop) possesses 
the nonrelativistic velocities, provided the elastic waves 
travelling along the string have sufficiently long wavelengths.

Let us consider  finite mass corrections to
the magnetic helicity \cite{moffatt69} of the string configuration, with the
further goal of calculating the time derivative of this quantity. The purpose
of this study is twofold. First, the role of configurations with a
nonzero magnetic helicity is
intensively discussed \cite{cornwall,giovan} in the connection with 
processes at the epoch of the electroweak phase transition. The dynamics of
such configurations affects, in the view of the anomaly
equation, the dynamics of some fermionic charges, in particular, the baryon
and/or lepton numbers \cite{giovan}.
Second, a number of papers have appeared recently
\cite{aldinger,goldstein,kamien},
where the dynamics of the twisting and writhing numbers
(see below) of the curve
is discussed from the geometrical point of view. In the meantime,
the gauge string in the coupled Higgs and vector field system should be
 governed by the corresponding field equations. In this respect
it would be interesting to compare the results inferred from the field
equations to the results inferred from pure geometrical approach of Refs.
\cite{aldinger,goldstein,kamien}.

The representation of the magnetic helicity $h_A$ in terms of the space Fourier
components of the gradient of the singular phase of the Higgs field,
Eq. (\ref{eq25}), found in \cite{kozhev95b} is useful. One has
\begin{eqnarray}
h_A&=&\int d^3x{\bf A}\cdot(\mbox{\boldmath$\nabla$}\times{\bf A})=
\int{d^3k\over(2\pi)^3}
\frac{i{\bf k}\cdot\left[{\bf v}({\bf k},t)
\times{\bf v}^{\ast}({\bf k},t)\right]}{({\bf k}^2m^{-2}_V+1)^2}
=              \nonumber\\
& &\Biggl(\frac{2\pi}{g}\Biggr)^2\int\frac{d^3k}{(2\pi)^3}
\Biggl(\frac{m^2_V}{{\bf k}^2+m^2_V}\Biggr)^2
\sum_{a,b}
\oint\oint d\sigma_a d\sigma_b\exp[-i{\bf k}\cdot({\bf X}_a
-{\bf X}_b)]    \nonumber\\
& &\times i{\bf k}\cdot[{\bf X}^\prime_a\times{\bf X}^\prime_b]/{\bf k}^2,
\label{hel}                                                     %(12)
\end{eqnarray}
which is nonzero only for 
configurations that are not invariant under the space inversion.
The shorthand notation ${\bf X}_{a,b}\equiv{\bf X}_{a,b}(\sigma_{a,b},t)$
is used hereafter.
Terms with $a\not=b$, after the momentum integration, give the
linking number
\begin{eqnarray*}
L[a,b]={1\over4\pi}\oint d\sigma_a\oint d\sigma_b
{{\bf X}_{ab}\cdot[{\bf X}^\prime_a\times{\bf X}^\prime_b]
\over|{\bf X}_{ab}|^3}
\end{eqnarray*}
of two contours \cite{frankk,moffatt92},
with the corrections suppressed exponentially as
$\exp(-m_V|{\bf X}_a-{\bf X}_b|)$.
The contribution of the typical term with $a=b$, after  momentum
integration, reads
\begin{eqnarray}
h_A(a=b)&\propto& W[a]-{1\over4\pi}\oint d\sigma_1\oint d\sigma_2
{{\bf X}_{12}\cdot[{\bf X}^\prime_1\times{\bf X}^\prime_2]\over
|{\bf X}_{12}|^3}
\left(1+m_V|{\bf X}_{12}|+{1\over2}m^2_V|{\bf X}_{12}|^2
\right)            \nonumber\\
& &\exp(-m_V|{\bf X}_{12}|),
\label{ab}
\end{eqnarray}
where ${\bf X}_{12}\equiv{\bf X}_a(\sigma_1)-{\bf X}_a(\sigma_2)$ refers
to the same contour $a$, and
\begin{eqnarray*}
W[a]={1\over4\pi}\oint d\sigma_1\oint d\sigma_2
{{\bf X}_{12}\cdot[{\bf X}^\prime_1\times{\bf X}^\prime_2]
\over|{\bf X}_{12}|^3}
\end{eqnarray*}
is the writhing number of the contour $a$ \cite{fuller}. The
$m_V$-dependent term in Eq. (\ref{ab}) is evaluated with the help of the
expansion Eq. (\ref{eqcl}) to give
$$\delta h_A(a=b, {\rm mass\mbox{ }correction})
\propto-{1\over2\pi m^2_V}\oint d\sigma {\bf X}^\prime_a\cdot
[{\bf X}^{\prime\prime}_a\times{\bf X}^{\prime\prime\prime}_a].$$
In the case of sufficiently smooth contours the latter can be represented as
$-T[a]/(m_VR)^2$, where
$$T[a]={1\over2\pi}\oint d\sigma{\bf X}^\prime\cdot[{\bf n}\times
{\bf n}^\prime]$$
is the twisting number \cite{frankk,moffatt92,fuller} of the contour $a$
whose normal vector
is ${\bf n}$ and the radius of curvature is $R$.
Thus, the part of the twist contribution to the helicity originating from
the finite width of the vortex is suppressed as
$(Rm_V)^{-2}$,
and the resulting expression for the helicity can be
written as \cite{sato95,kozhev95b}
\begin{equation}
h_A=\Phi_0^2\left\{\sum_{a}W[a]+2\sum_{a<b}L[a,b]\right\}.
\label{dop3}
\end{equation}

The rate of the helicity change in the course of the contour evolution
can be evaluated explicitly by taking the time derivative of the right hand
side of Eq. (\ref{hel}).  With the help of the relation
$${\partial\over\partial t}\oint d\sigma[{\bf k}\times{\bf X}^\prime]
\exp(-i{\bf k}\cdot{\bf X})=
i\oint d\sigma{\bf k}\times({\bf k}\times
[{\bf\dot X}\times{\bf X}^\prime])\exp(-i{\bf k}\cdot{\bf X}),$$
which can be verified by a straightforward calculation, one finds
\begin{eqnarray}
\dot h_A&=&\Phi_0^2
\int{d^3k\over(2\pi)^3}\left({m^2_V\over{\bf k}^2+m^2_V}\right)^2\sum_{ab}
\oint d\sigma_a\oint d\sigma_b
({\bf\dot X}_a-{\bf\dot X}_b)[{\bf X}^\prime_a
\times {\bf X}^\prime_b]         \nonumber\\
& &\times\exp(-i{\bf k}\cdot{\bf X}_{ab})   \nonumber\\
& &={\Phi_0^2m^3_V\over 8\pi}\sum_{ab}
\oint d\sigma_a\oint d\sigma_b
({\bf\dot X}_a-{\bf\dot X}_b)[{\bf X}^\prime_a
\times {\bf X}^\prime_b]\exp(-m_V|{\bf X}_{ab}|).
\label{cont}
\end{eqnarray}
It is clear that  terms with $a\not=b$ give an exponentially small
correction $\propto\exp(-m_V|{\bf X}_a-{\bf X}_b|)$. This is natural,
since the analogous terms in the expression
for $h_A$ give a contribution to the linking number $L[a,b]$ known to be
the topological invariant. The contribution of the terms with $a=b$ is
calculated with the help of the expansion  (\ref{eqcl})
taken at $t_1=t_2$ and $\sigma_2=\sigma_a+z$. One obtains
\begin{eqnarray}
\dot h_A&=&{\Phi_0^2m^3_V\over 8\pi}\sum_{a}\oint d\sigma_a
{\bf\dot X^\prime}_a\cdot[{\bf X}^\prime_a
\times{\bf X}^{\prime\prime}_a]
\int_{-\infty}^{+\infty}dz(-z^2)\exp(-m_V|z|)    \nonumber\\
& &={\Phi_0^2\over 2\pi}
\sum_a\oint d\sigma_a{\bf\dot X}_a\cdot[{\bf X}^\prime_a
\times{\bf X}^{\prime\prime\prime}_a].
\label{deriv}
\end{eqnarray}
The contribution to the time derivative turns out to be independent of $m_V$.
Note that the time derivative of helicity calculated in Eq. (\ref{deriv})
originates from the intrinsic dynamical contour motion. Indeed, in the case of
translational motion with 
constant velocity one can show, with the help of the Frenet equations, that
the right hand side of Eq. (\ref{deriv}) vanishes.

The above change of helicity is
identified as due to the change of the writhing number. Hence,
Eq. (\ref{deriv}), after dividing by $\Phi_0^2$, gives the time derivative
of the writhing number; see Eq. (\ref{dop3}). 
The presented derivation can be compared  to
the earlier derivation  \cite{aldinger} obtained by using  geometrical 
means. The equation found here coincides exactly with the equation obtained
in Ref. \cite{aldinger}.

One can  write Eq. (\ref{deriv}) in terms
of an infinitesimal deformation of the contour $\delta{\bf X}$ as
\begin{equation}
\delta W[a]=\frac{1}{2\pi}\oint d\sigma\delta{\bf X}\cdot
[{\bf X}^\prime\times{\bf X}^{\prime\prime\prime}].
\label{dwr}
\end{equation}
The transversal deformation preserving the gauge condition
${\bf X}^{\prime2}=1$ should look like $\delta{\bf X}=\delta X_b
{\bf b}$, where ${\bf b}$ is the unit vector of binormal. This can be
verified with the help of the Frenet equations. After an extensive use
of these equations, one can write Eq. (\ref{dwr}) in the form
$$\delta W[a]=-\frac{1}{2\pi}\oint d\sigma\kappa\delta X_b^\prime,$$
where $\kappa$ is the curvature of the contour. On the other hand, an
infinitesimal variation of the twisting number can be written as the
following chain of equations:
\begin{eqnarray}
\delta T[a]&=&\frac{1}{2\pi}\delta\oint d\sigma\tau=
\frac{1}{2\pi}\oint d\sigma\delta({\bf b\cdot n}^\prime)=
\frac{1}{2\pi}\oint d\sigma\left\{(\delta{\bf b}\cdot{\bf n}^\prime)-
(\delta{\bf n}\cdot{\bf b}^\prime)\right\}   \nonumber\\
&&=\frac{1}{2\pi}\oint d\sigma\left\{\delta{\bf b}
(-\kappa{\bf X}^\prime+\tau{\bf b})+\tau{\bf n}\cdot\delta{\bf n}\right\}
=-\frac{1}{2\pi}\oint d\sigma\kappa({\bf X}^\prime\cdot\delta{\bf b}).
\label{dtw}
\end{eqnarray}
The orthonormality property of the three vectors ${\bf n}$, ${\bf b}$, and
${\bf X}^\prime$ is used. This property permits one to write finally
that
\begin{equation}
\delta T[a]=\frac{1}{2\pi}\oint d\sigma\kappa\delta X_b^\prime
=-\delta W[a].
\label{sl}
\end{equation}
Hence, the dynamical evolution of the vortex string in the Abelian Higgs
model obeys the conservation law $W[a]+T[a]=S[a]=const$, where
the integration constant $S[a]$ is known as the  self-linking number of a 
curve $a$, with the property being the topological invariant
\cite{sato95,frankk,moffatt92,fuller,pohl}. Hence topological
invariance is obtained here, in fact, from the field equations of the 
Abelian Higgs model, without the assumption of conserved helicity.
The present conclusion is important, because an earlier derivation
\cite{moffatt92} of the above topological conservation law was based on
the conservation of  helicity. The latter is valid only approximately,
in the limit of an infinite conductivity of the medium 
\cite{moffatt69,moffatt92,field}.

In conclusion, let us discuss the meaning of the results obtained here.
First, the time derivative of the writhing number is independent
of the mass of the vector boson and coincides with the expression
\cite{aldinger} found in a completely different situation. This signals 
the universal dynamics of the writhing number of the objects spreading from
cosmic strings to  polymer chains.
Second, the physical meaning of Eq. (\ref{deriv}) is the following.
The dynamics of the $W^\pm$ condensates \cite{cornwall} and the electroweak
plasma \cite{giovan} in the electroweak phase transition
is helicity preserving. The same situation takes place in
magnetohydrodynamics \cite{moffatt69,moffatt92,field}. In all these cases
the change of helicity is due to the finite conductance of a medium and,
consequently, is dissipative. Only in
this case does a nonzero  scalar product ${\bf E\cdot H}$ contribute to
the time derivative of the magnetic helicity, provided there are no
objects that are not invariant under  space inversion.
The situation when such objects are present is considered in the present work.
These are gauge strings with the parity-noninvariant contours evolving
in accordance with the field equations. 
The above scalar product is nonzero in this case.
Using Eqs. (\ref{eq23}) and (\ref{dop1}) one can show that Eq. (\ref{deriv})
can be alternatively obtained from a direct evaluation of the right hand
side of the equation
$\dot h_A=-2\int d^3x{\bf E\cdot H}$. This demonstrates that  magnetic
helicity can change even
in the case of an infinite plasma conductivity, since the
electric field strength has a nonzero projection onto the magnetic field
strength for the closed mirror-noninvariant string. The
change is dissipationless, and the sign of the right hand side of
Eq. (\ref{deriv}) can be arbitrary.

\acknowledgments
I am indebted to Dr. Renzo L.~Ricca for bringing some relevant references,
especially Ref. \cite{moffatt92}, to my attention.
The study of that paper has promoted a clarification of some points
in the present work.


\begin{references}
\bibitem{fn1}
As was shown in Ref. \cite{orland}, such an expansion is not, in general,
a good guide to the stability of the vortex. However, the condition
$m_H\gg m_V$ adopted in the present paper means that, in terms of Ref.
\cite{orland}, the penetration depth is much greater than the correlation
length. This is deep inside  the stability region as pointed out in Ref.
\cite{orland}.
\end{references}
\end{document}